\documentstyle[12pt,epsfig]{article}
\newlength{\dinwidth}
\newlength{\dinmargin}
\newcommand{\be}{\begin{equation}}
\newcommand{\ee}{\end{equation}}
\newcommand{\bea}{\begin{eqnarray}}
\newcommand{\eea}{\end{eqnarray}}
\newcommand{\gsim}{\mathrel{\mathop{\kern 0pt \rlap
  {\raise.2ex\hbox{$>$}}} \lower.9ex\hbox{\kern-.190em $\sim$}}}

\def\bbox{{\,\lower0.9pt\vbox{\hrule \hbox{\vrule height 0.2 cm
\hskip 0.2 cm \vrule height 0.2 cm}\hrule}\,}}
\newcommand{\dsl}{\pa \kern-0.5em /}
\renewcommand{\t}{\theta}
\newcommand{\T}{\Theta}

\catcode`@=11 \@addtoreset{equation}{section} \catcode `@=12
\renewcommand{\theequation}{\thesection.\arabic{equation}}

\def\* {&=&}

\def\N{{\rm N}}
\def\M{{\rm M}}
\newcommand{\nc}{\newcommand}
\newcommand{\rnc}{\renewcommand}

\makeatletter
\rnc{\theequation}{\thesection.\arabic{equation}}
\@addtoreset{equation}{section}
\makeatother

\nc{\fig}[3]{
\begin{figure}
\centerline{\epsfxsize=#1\epsfbox{#2.eps}}
\caption{#3. \label{#2}}
\end{figure}
}

\def\CK{{\cal K}}

\def\CO{{\cal O}}

\def\a{\alpha}

\def\d{{\rm d}}
\def\ep{\epsilon}

\def\l{\lambda}

\def\n{\nu}
\def\th{\theta}

\def\t{\tau}

\def\G{\Gamma}
\def\T{{\rm T}}

\def\half{\frac{1}{2}}

\def\goto{\rightarrow}

\def\det{{\rm det}}

\begin{document}
\thispagestyle{empty}
\addtocounter{page}{-1}
\begin{flushright}
SNUST-01-0902\\
KIAS-P01045\\
PUPT-2008\\
{\tt hep-th/0110215}
\end{flushright}
\vspace*{0.3cm}
\centerline{\Large \bf Interacting Open Wilson Lines}
\vskip0.45cm
\centerline{\Large \bf from}
\vskip0.45cm
\centerline{\Large \bf Noncommutative Field Theories~\footnote{
Work supported in part by the BK-21 Initiative in Physics, 
the KRF International Collaboration Grant, the KRF Grant 2001-015-DP0082, 
the KOSEF Interdisciplinary Research Grant 98-07-02-07-01-5, 
the KOSEF Leading Scientist Program, and the KOSEF Brain-Pool Program.}}
\vspace*{1.1cm} 
\centerline{\bf Youngjai Kiem ${}^{a,d}$, Sangmin Lee ${}^b$, 
Soo-Jong Rey ${}^c$, Haru-Tada Sato ${}^a$}
\vspace*{0.6cm}
\centerline{\it BK21 Physics Research Division \& Institute of Basic Science}
\vspace*{0.25cm}
\centerline{\it Sungkyunkwan University, Suwon 440-746 \rm KOREA ${}^a$}
\vspace*{0.4cm}
\centerline{\it School of Physics, Korea Institute for Advanced Study, 
Seoul 130-012 \rm KOREA ${}^b$}
\vspace*{0.4cm}
\centerline{\it School of Physics \& Center for Theoretical Physics}
\vspace*{0.25cm}
\centerline{\it Seoul National University, Seoul 151-747 \rm KOREA ${}^c$}
\vspace*{0.4cm}
\centerline{\it Physics Department, Princeton University, Princeton, NJ
08544 \rm USA ${}^d$} 
\vspace*{0.8cm}
\centerline{\tt ykiem, haru@newton.skku.ac.kr, \hskip0.3cm 
            sangmin@kias.re.kr, \hskip0.3cm sjrey@gravity.snu.ac.kr}
\vspace*{1cm}
\centerline{\bf abstract}
\vspace*{0.3cm}
In noncommutative field theories, it was known that one-loop effective 
action describes propagation of non-interacting open Wilson lines, obeying
the flying dipole's relation. We show that two-loop effective action 
describes cubic interaction among `closed string' states created by open 
Wilson line operators. Taking d-dimensional 
$\lambda [\Phi^3]_\star$-theory as the simplest setup, we compute nonplanar
contribution at low-energy and large noncommutativity limit. We find that 
the contribution is expressible in a remarkably simple cubic interaction 
involving scalar open Wilson lines only and nothing else. We show that the 
interaction is purely geometrical and noncommutative in nature, depending 
only on sizes of each open Wilson line. 
\vspace*{1.1cm}

\baselineskip=18pt
\newpage
%%%%%%%%%%%%%%%%%%%%%%%%%%%%%%%%%%%%%%%%%%%%%%%%%%%%%%%%%%%%%%%%%%%%%%%%%%%
\section{Introduction}
%%%%%%%%%%%%%%%%%%%%%%%%%%%%%%%%%%%%%%%%%%%%%%%%%%%%%%%%%%%%%%%%%%%%%%%%%%%
The most significant feature of generic noncommutative field theories
is phenomenon of the UV-IR mixing \cite{uvir}. Recently, it is asserted that 
open Wilson lines (OWLs) \cite{rey1, rey2, ghi} 
are responsible for the phenomenon \cite{rey}:
long-distance excitations described by the open Wilson lines correspond to 
noncommutative dipoles \cite{dipole, stringdipole} 
(direct analogs of the Mott excitons \cite{rey} in metal under 
a strong magnetic field), and accounts for the peculiar long-distance dynamics 
in noncommutative field theories. It also implies that the open Wilson lines 
ought to be ubiquitous to any noncommutative field theory, be it gauge 
invariant or not, or Poincar\'e invariant or not. A partial evidence for 
the ubiquity is provided by the spin-independence of 
the generalized $\star_\N$-product \cite{kps}. Based on the insight,
in \cite{krsy1, krsy2}, it was proven that the nonplanar part of the 
complete one-loop effective action of a noncommutative {\em scalar} field 
theory is expressible entirely in terms of open Wilson lines 
--- and nothing else --- in a remarkably simple form. 
This feature is in support of the conjecture \cite{rey} that, 
as in $s-t$ channel duality in open and closed string theories \cite{dasrey}, 
quantum dynamics of elementary field is described, at long distance, by 
classical dynamics of open Wilson line. This duality underlies the phenomenon
of UV-IR mixing, and is quantified by the following relation
between dipole moment and energy-momentum four-vectors:
\bea
\ell^m = \th^{mn} k_n,
\label{ell}
\eea
nicknamed as `flying dipole's relation' \cite{rey}.
\begin{figure}[htb]
\centerline{\epsfxsize=5cm\epsfbox{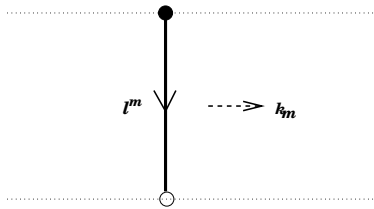}}
\caption{\sl Spacetime view of open Wilson line propagation. 
The noncommutativity is turned on the plane spanned by the two vectors, 
$\ell^m$ and $k_m$. 
\label{propagator}}
\end{figure}

The result of \cite{krsy1}\cite{krsy2} indicates that the one-loop effective 
action describes propagation of a non-interacting open Wilson line and takes 
schematically the form:
\bea
\Gamma_2 [W(\Phi)] = {1 \over 2} {\rm Tr}_{{\cal H}_{\rm dipole}} 
\Big( \widehat{W} \,\cdot \, {\cal K}_{-{d \over 2}} 
\, \cdot \, \widehat{W} \Big),
\label{oneloopresult}
\eea
where the scalar Wilson line operator is defined as
\bea
W_k [\Phi] = 
\int \d^d x {\cal P}_{\tau} \exp_\star \left( - g \int\limits_0^1 \d \tau 
\left\vert \dot{y}(\tau) \right\vert \Phi(x + y(\tau)) \right) 
\star e^{i k \cdot x},
\label{scalarowl}
\eea
${\cal H}_{\rm dipole}$ refers to
a one-`dipole' Hilbert space, and ${\cal K}_{-{d \over 2}}$ denotes spacetime 
propagation kernel of the dipole.

In this paper, we shall be extending earlier analysis to two-loop level,
and study \underline{interaction} among `closed string' states created by 
the open Wilson line operators. The main results are
(1) the two-loop effective action is expressible in terms of interaction
among the noncommutative dipoles created by the open Wilson lines, (2) 
interaction is cubic, and is entirely geometrical, as dictated by the 
flying dipole's relation, and (3) the interaction is suppressed at high 
energy-momentum by a nonanalytic damping factor. 
According to the flying dipole's relation, interaction among the 
noncommutative dipoles, if exists, ought to obey 
geometric constraints among the dipole moments. An outstanding question 
would then be building up an intuitive picture of the interaction, which we
try to answer in this work. 
\begin{figure}[htb]
\centerline{\epsfxsize=7cm\epsfbox{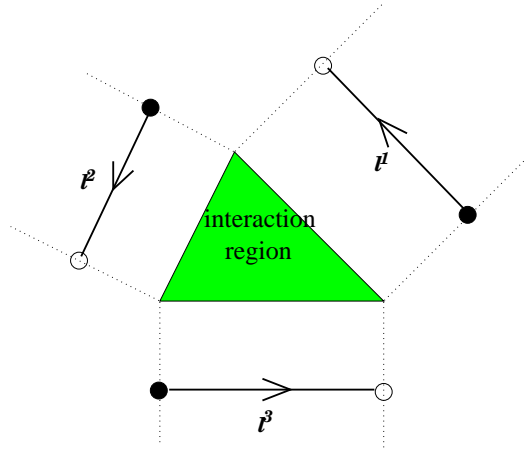}}
\caption{\sl Interaction of three noncommutative dipoles.
At interaction region, the dipoles are expected to interact locally pairwise.}
\end{figure}
The main results assert that cubic interaction among the open Wilson
lines is governed by a remarkably simple effective action, schematically
taking the following form:
\bea
\Gamma_3[W] = {\lambda_{\rm c} \over 3!} {\rm Tr}_{{\cal H}_{\rm dipole}} 
{\bf K}_3 \, 
\Big(
\widehat{W} \, \widehat{\star} \, \widehat{W} \, \widehat{\star} \, 
\widehat{W} 
\Big) 
\qquad {\rm where} \qquad
\lambda_{\rm c} = (\lambda \slash 2)^2.
\label{cubicint}
\eea
Here, ${\bf K}_3$ represents a weight-factor over ${\cal H}$, and 
the $\widehat{\star}$-product refers to a newly emergent noncommutative
algebra obeyed by the open Wilson lines as `closed strings'. 

This paper is organized as follows. In section 2, we rederive
the one-loop effective action \cite{krsy1, krsy2} via the saddle-point method.
In section 3, starting from two-loop Feynman diagrammatics, we obtain 
factorized expression of the two-loop effective action. In section 4, we 
evaluate the effective action via saddle-point method and show emergence of
snapped open Wilson lines. In section 5, we obtain the proclaimed result
Eq.(\ref{cubicint}).
%%%%%%%%%%%%%%%%%%%%%%%%%%%%%%%%%%%%%%%%%%%%%%%%%%%%%%%%%%%%%%
\section{One-loop Effective Action Revisited}
%%%%%%%%%%%%%%%%%%%%%%%%%%%%%%%%%%%%%%%%%%%%%%%%%%%%%%%%%%%%%%
\begin{figure}[htb]
\centerline{\epsfxsize=6cm\epsfbox{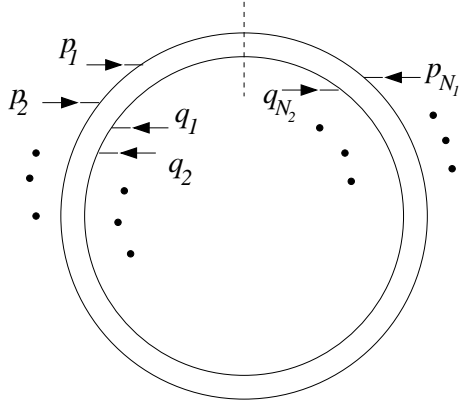}}
\caption{
\sl One-loop Feynman diagram for $\N$-point Green function.
There are $\N_1$ and $\N_2$-insertions of 
external momenta along the inner and outer boundary
of the Feynman diagram in double line notations.
\label{1lfeyn}}
\end{figure}

We first recapitulate aspects of the one-loop effective action \cite{krsy1, 
krsy2} relevant
for discussion in later sections. The effective action is defined as
\bea
\Gamma_{\rm 1-loop} = \sum_{\N=0}^\infty {1 \over \N!} \Gamma_\N,
\nonumber
\eea
viz. a sum over N-point, one-particle-irreducible Green function, $\G_\N$:
\bea
\G_\N [\{p_i\}, \{q_j\}] = \hbar \left(-\frac{\l}{2}\right)^\N
\sum_{\N_1+\N_2=\N} {C_{\{\N\}} \over (4 \pi)^{d/2}}
 \int\limits_0^{\infty} \frac{\d\T}{\T} 
\T^{-{d \over 2}+\N} \exp\left[-m^2 \T-\frac{\ell^2}{4\T} \right] 
J_{\N_1}(\ell) J_{\N_2}(-\ell).
\nonumber
\eea
Here, we have denoted N-dependent combinatoric factor as $C_{\{\N\}}$,
divided $\N$ external momenta into two groups: 
$\{p_1, \cdots p_{\N_1}\}$ and $\{ q_1, \cdots, q_{\N_2} \}$, and
defined $k = \sum_{i=1}^{\N_1} p_i = - \sum_{i=1}^{\N_2} q_i$, 
and $\ell := \theta \cdot k$ (consistent with the flying dipole's
relation). We have also defined a kernel $J_{\N_1}(\ell)$ by
\bea
J_{\N_1} (\ell,\{p_i\}) := 
\int_{0}^{1} \cdots \int_{0}^{1} \d \tau_1 \cdots \d \tau_{\N_1} \,
\exp\left[-i\sum_{i=1}^{\N_1} \tau_i p_i\cdot \ell - 
\frac{i}{2} \sum_{i<j=1}^{\N_1}\ep(\tau_{ij}) \, p_i\wedge p_j \right]
\label{starkernel}
\eea
in which $\tau_i \, \sim \tau_i +1$, ($i=1, 2, \cdots$) denote moduli 
parameter for $i$-th 
marked point around a boundary of the vacuum diagram, where a background 
field $\Phi$ with momentum $p_i$ is inserted, and 
$\tau_{ij} := (\tau_i-\tau_j)$. Similarly, $J_{\N_2} (- \ell, \{q_j\})$
is defined around the other boundary of the vacuum diagram.
Eq.(\ref{starkernel}) 
is precisely momentum-space representation of the $\star_\N$-product, 
originally introduced in \cite{garousi, trek2} \footnote {As $k \cdot \ell
= 0$, the kernel is invariant under a uniform shift of all moduli 
parameters: $\tau_i \rightarrow \tau_i + $(constant). Using the invariance,
one can shift the $\tau$-integral domain in Eq.(\ref{starkernel}) 
to $[-1/2, +1/2]$. This choice is preferred, as it displays Hermiticity 
property of $J_\N$ manifestly.}$\label{shift}$. 

In the large noncommutative limit, $\theta^{mn} \rightarrow \infty$ (compared 
to typical energy-momentum scale), the $\T$-moduli integral is evaluated 
accurately via the saddle-point method. Clearly, the saddle-point is 
located at $\T = |l|/2m$. Moreover, the leading-order correction is from
Gaussian integral over the quadratic variance in the exponent, and is easily
computed to be $(\pi \vert \ell \vert \slash m^3)^{1/2}$. Thus, putting all these pieces
together, the one-loop, N-point Green function is obtained as
\bea
\G_\N [\{p_i\}, \{q_j\}] &=& \hbar 
(4 \pi)^{-{d \over 2}}\left(\frac{|\ell|}{2m}\right)^{-{d \over 2}} 
\left( {2 \pi \over m \vert \ell \vert} \right)^{1/2} 
e^{-m|\ell|} \nonumber \\
&\times& \sum_{\N_1+ \N_2= \N} 
C_{\{\N\}}
\left\{ (- g)^{\N_1}|l|^{\N_1} J_{\N_1}(\ell) \right\}
\left\{ (- g)^{\N_2}|l|^{\N_2} J_{\N_2}(-\ell) \right\}, 
\label{factorized}
\eea
where $g := ( \l/4m)$. The factorized expression permits resummation
of the double-sum over $\N_1$ and $\N_2$.  
Indeed, taking carefully into account of the combinatorial factors $C_{\{\N\}}
={1 \over 2} (\N! \slash \N_1! \N_2!)$,  we find that each factor involving 
the $\star_\N$-kernel $J_\N$'s are exponentiated into the {\sl scalar} open 
Wilson lines. Convoluting the Green functions with the background 
$\Phi$-fields and summing over N, the result is 
\bea
\G_{\rm 1-loop} [W(\Phi)] &=& 
\frac{\hbar}{2} 
\int \frac{\d^d k}{(2\pi)^d} W_k [\Phi] \, \CK_{-{d \over 2}}(|\ell|) \, 
W_{-k}[\Phi].
\nonumber
\eea
Here,
\bea
\CK_{-{d\over2}} (\vert \ell \vert) = 
\left(2\pi {\vert \ell \vert \over  m } \right)^{-{d \over 2}} 
\left({2 \pi \over m \vert \ell \vert}\right)^{1/2} e^{-m\vert \ell \vert}
\nonumber 
\eea
denotes the propagation kernel of the noncommutative dipole, accounting  
for the UV-IR mixing and infrared singularity at $\ell \rightarrow 0$, and
\bea
W_k[\Phi] &=& \sum_{\N=0}^\infty
{1 \over \N!} \Big(-g \vert \ell \vert \Big)^\N
\int {\d^d p_1 \over (2 \pi)^d}
\cdots \int {\d^d p_\N \over (2 \pi)^d} 
(2 \pi)^d \delta^d \left( p_1 +\cdots + p_\N - k \right)
\nonumber \\
&& \hskip2.5cm 
\times \Big[ J_\N \left(\ell, \{p_i\} \right) \cdot 
\widetilde{\Phi}(p_1) \cdots \widetilde{\Phi}(p_\N) \Big]
\label{owldef}
\eea
denotes the {\sl scalar} open Wilson line operator, expressed as a convolution
of the $\star_\N$-product kernel $J_\N$, Eq.(\ref{starkernel}) \cite{krsy1, 
trek2, wise}. 

Note that, inferred from the general expression, Eq.(\ref{scalarowl}), 
saddle-point contour of the open Wilson line Eq.(\ref{owldef}) is a straight 
line: $y^m(\tau) = \theta^{mn} k_n \tau$, thus mapping compact $\tau$-moduli 
space 
(of topology ${\bf S}^1$, corresponding to each of the two boundaries 
in Fig.(\ref{1lfeyn}) to a straight line of interval $\ell$ in spacetime.  
Obviously, such straight open Wilson
lines obey the `flying dipole relation' Eq.(\ref{ell}). The contour will
begin to fluctuate, however, once contributions beyond the saddle-point are
taken into account, yielding plethora of higher-mode excitations.
In the rest of this paper, we will see that, at higher-loops, even the
saddle-point contour turns out different from a straight line. 
%%%%%%%%%%%%%%%%%%%%%%%%%%%%%%%%%%%%%%%%%%%%%%%%%%%%%%%%%%%%%%%%%%%
\section{Two-Loop Effective Action}
%%%%%%%%%%%%%%%%%%%%%%%%%%%%%%%%%%%%%%%%%%%%%%%%%%%%%%%%%%%%%%%%%%%
\subsection{Two-Loop Green Functions}
%%%%%%%%%%%%%%%%%%%%%%%%%%%%%%%%%%%%%%%%%%%%%%%%%%%%%%%%%%%%%%%%%%%
Begin with the two-loop nonplanar contribution to the N-point 
one-particle-irreducible Green functions. The general expression of these
Green functions have been obtained recently in \cite{master} using the 
worldline formulation approach \footnote{For recent study of multiloop 
$\Phi^3$-theory amplitudes from string theory, see \cite{rolandsato}.}. 
The result agrees with that obtainable
from the noncommutative Feynman diagrammatics, which we shall be presenting 
in Appendix A, and is given by:
\be\label{GPN}
\G_\N \left(\{p_a^{(a)}\} \right) 
= {\hbar^2 \lambda^2\over 24} \left( -{\lambda \over 2} \right)^{\N} 
\sum_{\N_1,\N_2,\N_3=0}^\N \sum_{\{\nu\}}
(2\pi)^d \delta \left(\sum_{a=1}^3\sum_{n=1}^{\N_a}p_n^{(a)} \right)
\,C_{\{\N\}} \, \Gamma_{\nu }^{(\N_1, \N_2, \N_3)}. 
\ee
Here, $C_{\{\N\}}$ denotes a combinatoric factor (see section 4.2), and 
\bea
\label{master1}
\begin{array}{rcl}
\Gamma_{\nu}^{(\N_1, \N_2, \N_3)}
&=&
\displaystyle
\frac{1}{(4\pi)^d}
\int\limits_0^\infty \cdots \int\limits_0^\infty 
\d \T_1 \d \T_2 \d \T_3 \,  e^{-m^2 (\T_1 +\T_2 +\T_3)}
\Delta^{\frac{d}{2}}(\T) \left( \prod_{a=1}^3 \int\limits_0^{\T_a}
\prod_{i=1}^{\N_a} \d \tau^{(a)}_i \right)
\\
&\times&
\displaystyle
\exp \left[  \frac{1}{2} \sum_{i,j=1}^\N p_i \cdot G^{(0)} \cdot p_j\right]
\exp \left[- \frac{1}{4}\Delta(\T) \sum_{a=1}^{3} \T_a k_a \circ k_a \right]
\\
&\times&
\displaystyle
\exp \left[ - \frac{i}{2} \sum_{a=1}^3 \left(
 P_a^- \wedge P_a^+ + \frac{1}{3} P_a \wedge P_{a+1} \right)
\right]
\\
&\times&
\displaystyle
\exp\left[\, -\frac{i}{4} \sum_{k<\ell}^{\N_a} 
\left(\nu^{(a)}_{k}+\nu^{(a)}_{\ell} \right)
\epsilon \left(\tau^{(a)}_{k\ell} \right) \, 
p^{(a)}_{k} \wedge p^{(a)}_{\ell} \right]
\\
&\times&
\displaystyle
\exp\left[  ~ i \Delta(\T) \sum_{a=1}^3 \T_a k_a \wedge
  \left( \sum_{i=1}^{\N_{a+2}} \tau_i^{(a+2)} p_i^{(a+2)}
 - \sum_i^{\N_{a+1}} \tau_i^{(a+1)} p_i^{(a+1)} \right)
 \right]  ~ ,
\end{array}
\eea
in which 
\bea
\Delta(\T) := \Big( \T_1 \T_2 + \T_2 \T_3 + \T_3 \T_1 \Big)^{-1}.
\label{Delta}
\eea
This is the main result, which we will take as the starting point for the 
analysis in later sections. 
In the rest of this subsection, we explain our notations in Eqs.(\ref{GPN},
\ref{master1}, \ref{Delta}). 
Derivation of Eq.(\ref{GPN}) via the noncommutative Feynman
diagrammatics will be relegated to Appendix A.

\begin{figure}[htb]
\centerline{\epsfxsize=10cm \epsfbox{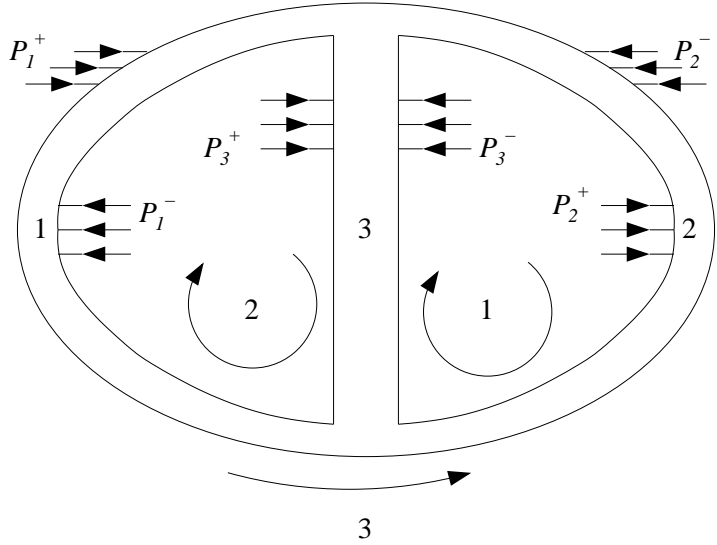}}
\caption{
\sl Two-loop Feynman diagram for N-point Green function. Note that we have 
labelled the legs and the vacuum diagram boundaries {\em dual} to each other.
The external momenta $p_i^{(a)}$, Feynman-Schwinger moduli parameters 
$\tau_i^{(a)}$,
and Moyal's phase-factor signs $\n_i^{(a)}$ are ordered from top to bottom, 
viz., $p_1^{(a)}$ refers the closest interaction vertex to the top, then 
$p_2^{(a)}$, and so on.
With this ordering convention, the $\tau's$ are ranging over 
$0 < \tau_\N < \cdots < \tau_1 < \T$ for each connected side of the three 
internal propagators.
\label{setup}}
\end{figure}

The Feynman diagram under consideration is depicted in Fig.4. 
We construct the {\sl planar} two-loop vacuum diagram
\footnote{At two loop and beyond, vacuum diagrams are classifiable into a 
planar diagram and the rest, 
nonplanar diagrams. If the number of twist insertion is zero, the vacuum
diagram is referred as planar. All other vacuum diagrams, with at least one 
insertion of the twist, are nonplanar ones. At one loop, by default, the 
vacuum diagram is planar.} by Wick-contracting two 
$\lambda[\Phi^3]_\star$-interaction vertices via three
internal propagators. Label the propagators as $a=1,2,3$ and draw them
in double-line notation, as is natural for noncommutative field theories. 
Moduli parameters $\T_1, \T_2, \T_3$ refer to the
Feynman-Schwinger parameters of the three internal propagators, and range
over the moduli space of two-loop vacuum Feynman diagram: 
${\cal M}_{\rm 2-loop} = [0, \infty) \otimes [0, \infty) \otimes [0, \infty)$. 
On the vacuum Feynman diagram, we mark N points at locations
$\tau^{(a)}_i$, where $a=1\pm,2\pm,3\pm$ and $i = 1, 2, \cdots, \N_a$, and 
insert background $\Phi$-fields. Each group of the $\Phi$-field insertions 
is classifiable into those affixed from the inner and the outer 
boundaries. Sum
over all possible insertion of the external lines is given by integration over
the moduli parameters $\tau_i^{(a)}$'s over $[0, \T_a]$, and $\tau_{ij}^{(a)}$
refers to $\left(\tau_i^{(a)} - \tau_j^{(a)}\right)$.

Momenta of background $\Phi$-fields attached at $a$-th internal propagator 
are labelled
as $\{ p_i^{(a)}\}$, where $a = 1,2,3$ and $i = 1, 2, \cdots, \N_a$. 
Total momentum injected on the $a$-th internal propagator is denoted as:
\bea
P_a^\pm \equiv \sum_{i=1}^{\N_a} \frac{1 \pm \n_i^{(a)} }{2} p_i^{(a)}
\qquad {\rm and} \qquad
P_a \equiv P_a^+ + P_a^- ~,
\label{ftmm}
\eea
where $\pm$ refers to the `left' and the `right' sides of $a$-th boundary
(see Fig.\ref{setup}), respectively,
and $\n_i^{(a)}$ takes $\pm1$ depending on whether the background
$\Phi$-field is attached from the `left' or the `right' side of 
the double-lined internal propagator. 
We have also introduced the total momentum inserted on each {\em worldsheet 
boundary} via
\bea
k_{a+2} = P^+_{a} + P^-_{a+1} ~ .
\label{bdrymm}
\eea

Various products in Eq.(\ref{master1}) are defined as follows:
\bea
A \cdot B := \delta^{mn} A_m B_n,  \qquad
A \wedge B := \theta^{mn} A_m B_n, \qquad
A \circ B := \delta_{mn} (\theta \cdot A)^m (\theta \cdot B)^n.
\nonumber
\eea

In this paper, following \cite{krsy1}, we work exclusively in the limit of
\underline{low energy-momentum}, \underline{large noncommutativity}, and 
\underline{weak external field}: 
\bea
p_i^{(a)}/m = {\cal O}(\epsilon^{+1}),\quad
m^2 \theta^{mn} = {\cal O} (\epsilon^{-2}),  \quad 
{\rm and} \quad
\l{\Phi}/m^2  = \CO (\ep^{+1}) \qquad {\rm as} \qquad
\epsilon \rightarrow 0^+. 
\label{ourlimit}
\eea
The first two limits ensure that various energy-momentum products in the 
exponent of Eq.(\ref{master1}) are hierarchically separated as
\bea
p_i^{(a)} \cdot p_j^{(b)} \T \quad \ll \quad p_i^{(a)} \wedge p_j^{(b)} 
\quad \ll \quad  m^2 \T \quad \sim \quad p_i^{(a)} \circ p_j^{(b)} \slash \T, 
\label{hierarchy}
\eea
at the saddle-point value, $\T \sim \ell/2m$, we will eventually find. 
As shown in \cite{krsy1, krsy2} and reviewed in section 2, 
the limit Eq.(\ref{ourlimit}) and the relation Eq.(\ref{hierarchy}) 
simplified computation of the one-loop effective action enormously.
The third limit of Eq.(\ref{ourlimit}), viz. the weak external field limit,
also simplifies the moduli-space integral. Schematically, the integral is
given by 
\bea
\int\limits_0^\infty \frac{\d \T}{\T} \T^{-{d \over 2}} 
\exp\left(-m^2 \T-\frac{\ell^2}{4\T}\right) 
\sum_{\N=0}^\infty \frac{1}{\N!} \T^{\N} (-\l \Phi)^\N ~ ,
\nonumber
\eea
and appears that the saddle-point approximation
might break down for the summand of large $\N$.  The sum over $\N$, however,
is estimated ${\cal O}\left(e^{-\T \l \Phi}\right)$, and hence is
negligible once the third limit in Eq.(\ref{ourlimit}) is taken. 
We note that the exponent of the open Wilson 
line ($\sim e^{-\l \ell \Phi /m}$) still remains of ${\cal O}(1)$ 
in the limit Eq.(\ref{ourlimit}). The same limit Eq.(\ref{ourlimit}) 
will be taken in computing the two-loop effective action, and will turn out
to yield considerable simplification, far more remarkable than in the 
one-loop computation.

For later convenience, we refer to the last three exponential terms 
in Eq.(\ref{master1}), involving $\wedge$-product, 
as $\Xi_1, \Xi_2, \Xi_3$, respectively. At this point, it suffices to
note that $\Xi_1$ is {\sl independent} of the moduli parameters 
$\tau_i^{(a)}$'s, $\Xi_2$ depends on the {\sl ordering} of the moduli 
parameters, and the exponent of $\Xi_3$ depends {\sl linearly} on the 
moduli parameters.   

%%%%%%%%%%%%%%%%%%%%%%%%%%%%%%%%%%%%%%%%%%%%%%%%%%%%%%%%%%%%%%%%%%%%%%%%
\subsection{Factorization}
%%%%%%%%%%%%%%%%%%%%%%%%%%%%%%%%%%%%%%%%%%%%%%%%%%%%%%%%%%%%%%%%%%%%%%%%
As is well-known, the one-particle-irreducible diagrams in field theories 
are derivable from connected diagrams in string theory (see \cite{kiemlee}
for an example in the two-loop context). The open string theory 
diagrams are labelled by external momenta of vertex operators inserted along 
the worldsheet
boundaries, with a fixed cyclic ordering on each boundary, but are otherwise
insensitive to details of the momentum distribution on a given boundary.
On the other hand, according to the steps leading to Eq.(\ref{master1}), 
the field theory diagrammatics appear to distinguish, in Fig.(\ref{setup}),
the different parts of a given boundary. Specifically, whereas the worldsheet 
boundary momenta are $k_a$'s, the Feynman diagrammatics
Eq.(\ref{master1}) is expressed in terms of individual $P^\pm_a$'s. See
Eq.(\ref{bdrymm}). 

Can we reorganize a collection of Feynman diagrams belonging to the same
graph-theoretic combinatorics and simplify them so that the correspondence 
to the string theory diagram becomes manifest? 
The answer is affirmatively \underline{positive}. 
Steps leading to the answer are as follows. First, both the local phase-factor in $\Xi_2$ and 
the $\tau$-dependent phase-factor in $\Xi_3$ are readily splitted into
three parts, each of which is attributable to the three boundaries, as 
depicted in Fig.(\ref{setup}). 
Second, utilizing the overall energy-momentum conservation 
$P_1 + P_2 + P_3 = 0$, the $\tau$-independent 
phase-factor in $\Xi_1$ is re-expressible as:
\bea
\Xi_1 = \exp \left[ - \frac{i}{2} \sum_{a=1}^3 \left(
 P_a^+ \wedge P_{a+1}^- + \frac{1}{3} k_a \wedge k_{a+1} \right)
\right] ~ .
\nonumber
\eea
The first term consists of $\wedge$-product among partial momenta carried
by the two halves of each boundary, whereas the second term depends only 
on the net momenta carried by each boundary. Reorganizing this way, 
the last three exponentials in Eq.(\ref{master1}) are factorizable 
in the following way: 
\bea
\Xi_1 \Xi_2 \Xi_3
&=&
\prod_{a=1}^3 \Big( \Xi_1^a \Xi_2^a \Xi_3^a \Big) \times 
\exp\left[ -\frac{i}{2}\sum_{a=1}^{3} \frac{1}{3}k_a \wedge k_{a+1} \right]
~,
\label{decomp}
\\
\Xi_1^a
&=&
\exp \left[ -\frac{i}{2} P^+_{a+1} \wedge P_{a+2}^- \right] ~,
\label{xi1}
\\
\Xi_2^a
&=&
\exp\left[
-\frac{i}{2} \Big( \sum_{i<j} p_i \wedge p_j \ep(\tau_{ij})\Big)^{(a+1)+}
+\frac{i}{2} \Big( \sum_{i<j} p_i \wedge p_j \ep(\tau_{ij})\Big)^{(a+2)-}
\right] ~,
\label{xi2}\\
\Xi_3^a
&=&
\exp\left[ -i\Delta(\T) \left(\sum \tau p \right)^{(a+1)+} \wedge
\left( \T_{a+2} k_{a+2} - \T_a k_a \right) \right]
\nonumber \\
&\times&
\exp\left[ - i\Delta(\T) \left(\sum \tau p \right)^{(a+2)-} \wedge
\left( \T_a k_a  - \T_{a+1}k_{a+1}\right) \right] ~.
\label{xi3}
\eea

A remark is in order. Consider, for instance, along the $(a=3)$ boundary, 
making a move of an external $\Phi$-field line from the $(1+)$ side 
to the $(2-)$ side while preserving the cyclic ordering. Clearly, each 
component of the exponentials $\Xi^3_{1,2,3}$ makes a jump when the line  
crosses the borders between $(1+)$ and $(2-)$ sides.
It turns out that, for both for the top and the bottom borders, the 
jump cancels out. That is, 
for all cyclically ordered background $\Phi$-field insertions along each        
boundary, the phase-factors $\Xi_1 \Xi_2 \Xi_3$ are continuous and periodic
as the insertion moduli $\tau$'s are varied.
This is, of course, what one expects from underlying string theory
diagrams.

We now rescale the moduli parameters as $\tau^{(a\pm)} \goto  
\mp \tau^{(a\pm)} \T_a$. 
Then, $\tau$ runs over $[-1, 0]$ for the points on the $(+)$ side, 
and over $[0, +1]$ for the points on the $(-)$ side,
where the two sides belong to the same worldsheet boundary.
The sign flip on the $(+)$ side amounts to aligning 
the two integration regions in the same direction 
around the oriented worldsheet boundary: $ \int_0^1 d\tau^{(a+)}
 \rightarrow \int_{-1}^0 d\tau^{(a+)}$ renders the $\tau$-moduli
increment on the $(+)$ side
coincide with the worldsheet boundary direction 
shown in Fig.(\ref{setup}).
Suppose the background $\Phi$-fields are inserted $\N_1^+$ and $\N_2^-$ 
times on the $(1+)$ and $(2-)$ sides of the $(a=3)$ boundary, respectively. 
After the aforementioned redefinition of $\tau$'s,
the $\Xi^3_1$ and $\Xi^3_2$ are combined to
\footnote{
Precisely the same construction goes through for $a=1,2$ boundaries, and
the derivation will be omitted.
}
\bea
\label{ff1}
\Xi_1^3 \Xi_2^3 =
\exp\left[
+\frac{i}{2} \sum_{i<j=1}^{\N_1^+ + \N_2^-} p_i \wedge p_j \ep(\tau_{ij}) 
\right] ~ .
\eea
Furthermore, $\Xi_3^3$ is simplifiable considerably as
\be
\label{ff2}
\Xi_3^3 =
\exp\left[ -i\sum_{i=1}^{\N_1^+} (\tau p_i)^{(1^+)} \cdot (-\a_1) 
- i \sum^{\N_2^-}_{i=1} (\tau p_i)^{(2^-)} \cdot (+\a_2) \right] ~ .
\ee
Here, $\alpha$'s are vectors formed out of the dipole moments $\{\ell_a\}$ via
\bea
\a_1 &=& t_1 ( t_2 l_2 - t_3 l_3)~ , \qquad
\a_2 = t_2 ( t_3 l_3 - t_1 l_1)~ ,   \qquad
\a_3 = t_3 ( t_1 l_1 - t_2 l_2)~ .
\label{alphas}
\eea
\bea
t_{a} &=& \sqrt{\Delta} \T_a  \qquad {\rm where} \qquad 
(t_1t_2 + t_2t_3 + t_3t_1) = 1.
\label{ts}
\eea
The overall energy-momentum conservation, $k_1 + k_2 + k_3 = 0$, puts 
the three dipole vectors $\{\ell_a\}$ form a triangle, referred as
`dipole-triangle'. Geometrically, for non-negative values of $\{t_a\}$, 
the vectors $\{\a_a\}$ in Eq.(\ref{alphas}) split the `dipole-triangle' 
into three pieces (See Fig. 5), viz.
\bea  
l_1 = \alpha_3 - \alpha_2 , 
\qquad 
l_2 = \alpha_1 - \alpha_3 , 
\qquad
l_3 = \alpha_2 - \alpha_1.
\label{elldef}
\eea

\begin{figure}[htb]
\centerline{\epsfxsize=8cm\epsfbox{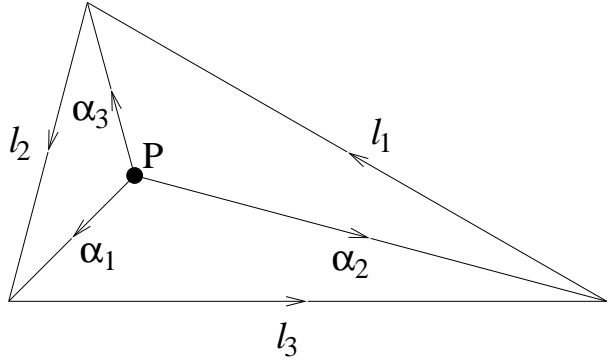}}
\caption{\sl
Due to momentum conservation ($k_1+k_2+k_3 = 0$), the vectors $l_1,l_2,l_3$ 
form a triangle. For any given value of the 
normalized moduli $\{t_a\}$, the $\a$'s split the triangle 
into three pieces. The saddle point conditions in Section 4 
demand that the angle between a pair of $\a$'s 
is $2\pi/3$.}
\label{alpha} 
\end{figure}

\noindent
Observe now that, after the factorization, exponentials in 
Eq.(\ref{ff1}) and Eq.(\ref{ff2}) are altogether strikingly reminiscent 
of the $\star_\N$-product kernel, Eq.(\ref{starkernel}), encountered at 
one-loop computation! 
As we will see shortly, they are resummable into a scalar open Wilson line
whose contour is not a straight line but a snapped wedge. Thus,  
introduce the two-loop $\star_\N$-product kernel as \footnote{
How they are related to the $\star_\N$-products and the
scalar open Wilson lines will be elaborated in the  next section.}
\be
\label{jhat}
\widehat{J}_{(\N_1^+, \N_2^-)}(-\a_1,\a_2) =  
        \prod_{i=1}^{\N_1^+} \int\limits_{-1}^0    \d \tau^{(1^+)}_i
        \prod_{j=1}^{\N_2^-} \int\limits_0^{+1} \d \tau^{(2^-)}_j ~
  \Xi^{(3)}_1 \Xi^{(3)}_2 \Xi^{(3)}_3 ~  \qquad
{\rm for} \qquad a=3 ,
\ee
and, similarly, $\widehat{J}_{(\N_2^+, \N_3^-)} ( -\a_2, + \a_3)$,
$\widehat{J}_{(\N_3^+, \N_1^-)} (-\a_3, +\a_1)$ for $a=$1 and 2, respectively.

We have denoted the two-loop kernels $\widehat{J}$'s hatted in order to 
emphasize functional difference of them from the one-loop kernel $J$'s. 
Nevertheless, the two are intimately related each other. To see this, use 
the identity $t_1t_2 +t_2t_3+t_3t_1 = 1$, and split the exponential term 
in Eq.(\ref{decomp}) into three parts: 
\bea
\exp\left[-\frac{i}{2} t_2t_3 k_2\wedge k_3 \right]
\exp\left[-\frac{i}{2} t_3t_1 k_3\wedge k_1 \right]
\exp\left[-\frac{i}{2} t_1t_2 k_1\wedge k_2 \right].
\nonumber
\eea
Permutation symmetry among the three exponentials suggests that each 
exponential is attachable to the three $\hat{J}$'s for $a=1,2,3$.
Indeed, using the identity
\bea
P_1^+ \wedge P_2^- + t_1t_2 k_1 \wedge k_2 +
P_1^+\cdot \a_1 + P_2^- \cdot \a_2 = 0,
\nonumber
\eea
derivable from the overall energy-momentum conservation,
one can show readily that the $\widehat{J}$'s in Eq.(\ref{jhat}) satisfies 
the following decomposition rule:
\bea
\widehat{J}_{(\N_1^+, \N_2^-)}(-\a_1,\a_2)  
= 
\exp\left[\frac{i}{2} t_1t_2 k_1 \wedge k_2\right ]
{J}_{\N_1^+}(-\a_1) {J}_{\N_2^-}(\a_2) ,  
\label{decomposition}
\eea
where $J$'s are to be understood as Eq.(\ref{starkernel}) but with 
the domain of moduli integrals shifted to $[-1/2, +1/2]$. 
Thus, the two-loop $\star_\N$-product 
kernel is decomposable into a product of two, one-loop 
$\star_\N$-product kernels. A notable feature of the decomposition is
that the product among the one-loop kernels is not an ordinary product,
but is a sort of (moduli-dependent) Moyal product. Moreover, decomposed
as in Eq.(\ref{decomposition}), the $\widehat{J}$'s display 
Hermiticity property manifestly --- 
with the integral domain shifted to $[-1/2,
+1/2]$, each of the two one-loop kernels in Eq.(\ref{decomposition}) is
manifestly Hermitian, and, under Hermitian conjugation and interchange
between $1 \leftrightarrow 2$, the newly emergent Moyal product on the
right-hand-side of Eq.(\ref{decomposition}) remains unchanged.
%%%%%%%%%%%%%%%%%%%%%%%%%%%%%%%%%%%%%%%%%%%%%%%%%%%%%%%%%%%
\section{Two-Loop Moduli Space Integral}
%%%%%%%%%%%%%%%%%%%%%%%%%%%%%%%%%%%%%%%%%%%%%%%%%%%%%%%%%%%

%%%%%%%%%%%%%%%%%%%%%%%%%%%%%%%%%%%%%%%%%%%%%%%%%%%%%%%%%%%%
\subsection{Saddle-Point Analysis}
%%%%%%%%%%%%%%%%%%%%%%%%%%%%%%%%%%%%%%%%%%%%%%%%%%%%%%%%%%%%
We now perform, in Eq.(\ref{master1}), the integration over 
$\T_1$, $\T_2$ and $\T_3$, viz. moduli space of the two-loop, 
planar, {\sl vacuum} Feynman diagram. 
Here enters the utility of the limit Eq.(\ref{ourlimit}), 
as the integration is simplified considerably. 
Specifically, (1) the low energy limit, $p^{(a)}_i \cdot p^{(b)}_j \ll m^2$, 
permits to drop the first exponential in Eq.(\ref{master1}), 
(2) the large noncommutativity, $m^2 \T \sim k_a \circ k_a \slash \T$,
permits to evaluate the integrals via the saddle-point method, and 
(3) the weak field limit, $\lambda \Phi \ll m^2$, permits to drop, 
in the saddle-point analysis, contribution of power-series in $(\lambda 
\Phi \T)$. We will find that, as a result of these simplifications, a 
version of open Wilson lines emerges naturally as the saddle-point 
configuration.   

We begin with detailed consideration of the moduli space integral.  
The integral is decomposable, via the change of variables Eq.(\ref{ts}), 
into one-dimensional integral over the `size' $\Delta(\T)$, and
two-dimensional integral over the `angles' $t_1, t_2, t_3$ (subject to the 
constraint $(t_1 t_2 + t_2 t_3 + t_3 t_1) = 1$). The decomposition
is suited the most for the saddle-point analysis, as the $\Xi_{1,2,3}^{(a)}$ 
are 
independent of the `size' $\Delta(\T)$, which is evident from Eqs.(\ref{ff1}, 
\ref{ff2}) and the definition of $\alpha$'s in Eq.(\ref{alphas}).
Domain of the moduli space in these variables is readily identifiable. 
Evidently, the `size' variable
ranges over $\Delta =[0, \infty)$. For the `angle' variables, 
using Eq.(\ref{alphas}), one can map their range to the range of
the $\alpha$-vectors. The latter is seen equal to the {\sl interior} of the 
`dipole-triangle' formed out of the three dipole moments, 
$\{\ell_a\}$: as the `angle' variables are varied, the junction 
point $P$ of the three $\alpha$ vectors sweeps out 
the `dipole-triangle' exactly once. See Fig.(\ref{alpha}). 

We now identify saddle-point value of $\Delta(\T)$ and 
saddle-point location of $P$, and compute the dominant contribution to the 
moduli space integral. Specifically, for the saddle-point analysis, take 
the following part of the integrand only:
\begin{equation}
 \exp F(\T_1 , \T_2 , \T_3 ) :=
  \exp \left[ - m^2 ( \T_1 + \T_2 + \T_3 )
   - \frac{1}{4} \Delta(\T)  (\T_1 l_1^2 + \T_2 l_2^2 +
      \T_3 l_3^2 ) \right] ~,
\label{sadstart}
\end{equation}
viz. the part dependent on the `size' variable. 
The saddle-point conditions, 
$\left( \partial F \slash \partial \T_a\right) = 0$, are then 
expressible, in terms of the `polar' variables, as 
\begin{equation}
 \Delta(\T) = \frac{4m^2}{L^2} 
\label{sizesad}
\end{equation}
and
\begin{equation}
 L \equiv | t_1 l_1 - t_2 l_2 |
 = |t_2 l_2 - t_3 l_3 | = |t_3 l_3 - t_1 l_1 |  ~.
\label{anglesad}
\end{equation}
Geometrically, the `size' condition Eq.(\ref{sizesad}) fixes $\Delta(\T)$ 
in terms of the area of the
`dipole-triangle' (which grows large in the large noncommutativity limit), 
while the `angle' condition
Eq.(\ref{anglesad}) puts the inner-angle between two adjacent $\a$'s to 
$(2\pi/3)$
\footnote{In Appendix A, we solve these saddle-point conditions
and determine explicitly the saddle-point value of $\{t_a\}$ in terms of 
the dipole moments $\{\ell_a\}$.}. The saddle-point is thus located at
\begin{equation}
\T_a = \Delta^{-1/2}(\T) t_a = \frac{L}{2m} t_a 
    = \frac{ |\alpha_a | }{2m} ~ ,
\label{wow}
\end{equation}
for which the saddle-point value of the exponent
\begin{equation}
 F ({\rm saddle-point}) = \left. - m \left( | \alpha_1 |
 + | \alpha_2 | +  |\alpha_3| \right)
   \right\vert_{\rm saddle~point} 
\label{leadingorder}
\end{equation}
is geometrically a linear sum of lengths of the three vectors, 
$\{\alpha_a\}$. 
As we will show in the next subsections, simple geometric description of
the saddle-point Eq.(\ref{wow}) permits the two-loop effective action 
expressible entirely in terms of the scalar open Wilson lines. 

We close the saddle-point analysis with a few remarks that warrant 
further technical elaboration. 
First, as in the one-loop computation in Section 2, 
one can perform the moduli integrals beyond the saddle-point, 
Eq.(\ref{leadingorder}). The leading-order correction is from 
quadratic variance of $F(\T)$ in the neighborhood of the saddle-point,
and, upon Gaussian integral, yields a pre-exponential correction, denoted
as $(\Delta_F)^{-3/2}$. In Appendix C, we compute $(\Delta_F)^{-3/2}$ 
and find that it also admits a simple geometric interpretation. 
Second, in setting up the saddle-point analysis, 
one might be concerned with potential contribution to Eq.(\ref{sadstart})  
from angular variation, viz. $t$-dependent, $\wedge$-product part. 
Fortuitously, in the large noncommutativity limit Eq.(\ref{ourlimit}), 
the angle-dependent part drops out of the analysis. 
To illustrate this, consider evaluating 
the moduli integrals over $(\Delta, t_1,t_2)$ via the saddle-point analysis.  
Recall that the $\wedge$-product part is independent of $\Delta$, and hence
does not affect the $\Delta$ integral. Thus, after the $\Delta$-integral,
we are left with
\bea
\int \d^2t e^{-mLf(t)} g(t),
\nonumber
\eea
where $f(t)$ refers to a dimensionless function of $t$, 
whose saddle-point $t^*$ is determined by Eq.(\ref{anglesad}). 
The function $g(t)$ includes the $\widehat{J}$ 
part (the $\wedge$-product part) and powers of $t$. 
From the expression, one readily finds that
the function $g(t)$ and its derivatives 
are of ${\cal O}(1)$.  Taylor-expanding $g(t)$ around $t^*$, 
a simple power-counting reveals that the $n$-th order 
term  is suppressed by a factor of $(mL)^{-n/2}$. 
Hence, in the limit $mL \gg 1$ we are working with, the function $g(t)$ does
not affect the saddle-point analysis based solely on the function 
$e^{-mLf(t)}$. 

\begin{figure}[htb]
\centerline{\epsfxsize=8cm\epsfbox{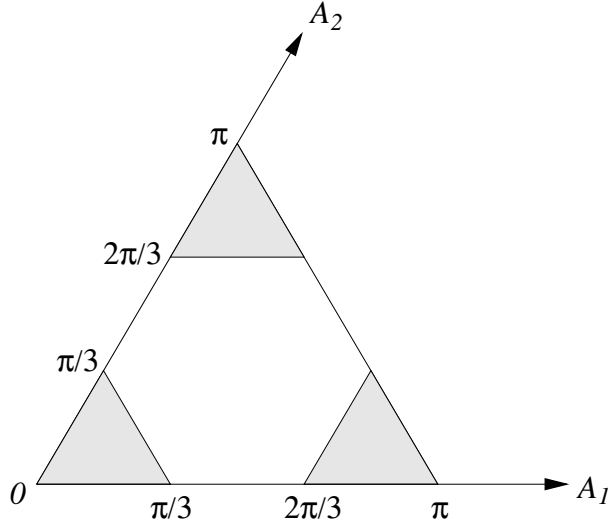}}
\caption{\sl Moduli space of the `dipole-triangle'.  A point in the 
moduli space represents a particular \underline{shape} 
of the dipole-triangle (defined up to size) whose two angles, $A_1, A_2$,
are the coordinates of the point.  Three corner
regions are shaded. 
\label{tri}}
\end{figure}

Third, a simple geometric consideration reveals that the 
`$(2\pi/3)$-condition', Eq.(\ref{anglesad}), 
is violated if any of the inner-angles in the `dipole-triangle' 
exceeds $2\pi/3$. 
Consider the moduli space of all possible `dipole-triangle' is
depicted in Fig.(\ref{tri}), where it is divided into a central region
and three corner regions.  The previous saddle-point
analysis is valid in the central region. In the corner regions, where
one of the three inner-angles in the `dipole-triangle' exceeds $2 \pi /3$,
the saddle-point analysis breaks down.  For instance, if the inner-angle 
between $l_1$ and $l_2$ is bigger than $2\pi/3$, 
the naive saddle-point analysis yields a negative value of $\T_3$
at the saddle-point, and hence the point $P$ lies outside 
the `dipole-triangle'.  As the moduli integrals cover only the 
positive values of $\T_a$, we will have to devise an another
approximation method. This turns out to be straightforward.  
In the lower-left corner region of Fig.(\ref{tri}), one can evaluate 
the $\T_1$ and $\T_2$ integrals, still utilizing the saddle-point method. 
The remaining $T_3$ integral is then schematically of the form:
\bea
\int \d \T_3 e^{-m^2\T_3} \T_3^{\N_3} \quad \sim \quad m^{-2\N_3} ~ ,
\nonumber
\eea
which eventually leads to the resummation 
\bea
\sum_{\N_3} \frac{1}{\N_3!} \left(-\frac{\l\Phi}{m^2}\right)^{\N_3} ~ .
\nonumber
\eea
Thanks to the weak field limit, $\l\Phi/m^2 \ll 1$ in Eq.(\ref{ourlimit}), 
we can safely keep the leading-order, $\N_3=0$ term only, viz. set the 
entire resummation to be unity. As this renders no insertion of the
background 
$\Phi$-field on the $a=3$ internal propagator, it sets $P_3^+ = P_3^- = 0$.  
It turns out that the rest of the saddle-point method is applicable
and the geometric structure does not change, except that now 
$\a_3 = 0$, $\a_2 = -\ell_1$ and $\a_1 = \ell_2$.
In the other two corner regions in Fig.(\ref{tri}), the same analysis 
holds regarding $\T_1, \T_2$ integrals and ensures no-insertion of the 
background $\Phi$-field on $a=1, 2$ internal propgators. 
A nothworthy point is that these three
corners of the moduli space correspond in the two-loop vacuum Feynman
diagrammatics to the degeneration limit where one of the three internal 
propagators is shrunken to a zero length. 
If the moduli space is extended to accomodate `topology change' 
across the three corners, the resulting diagrams are 
one-particle-{\sl reducible}, dumb-bell shaped Feynman diagram. 
Intriguingly, after a straightforward computation, we were able to show that, 
functionally, these one-particle-reducible diagrams are smoothly connected 
to the diagrams at the three corners as $\T_a$'s are extended to negative 
values. It implies that, within the saddle-point analysis,
one might take an alternative viewpoint that effective field theory
of noncommutative dipoles originate not just from one-particle-irreducible
diagrams but also one-particle-reducible ones. Needless to say, it is 
tantalizingly reminiscent of the worldsheet diagrammatics in string theory.  
%%%%%%%%%%%%%%%%%%%%%%%%%%%%%%%%%%%%%%%%%%%%%%%%%%%%%%%%%%%%%%%%%%%%%%%
\subsection{Combinatorics}
%%%%%%%%%%%%%%%%%%%%%%%%%%%%%%%%%%%%%%%%%%%%%%%%%%%%%%%%%%%%%%%%%%%%%%%
We next add up Feynman diagrams with different combinatorics of 
background $\Phi$-field insertion, and obtain the two-loop effective action.
In Eq.(\ref{master1}), the summations
\bea
 \sum_{\N_1+\N_2+\N_3=\N} \sum_{\{\nu\}} ~ 
\nonumber
\eea
are over all possible insertion of total N background $\Phi$-fields;
each $a$-th internal propagator carries N$_a$ out of them 
(with all possible orderings, as $\tau$'s run from $0$ to $1$ independently). 
The sign-factor $\nu_i^{(a)}$ keeps track of whether a given $\Phi$-field
is attached on the left or the right side of the propagator. 
In the commutative setup, the left/right insertions are not distinguished, 
and hence the summands consist of $3^\N$ combinatorically distinct diagrams. 
In the noncommutative case, the inclusion of the last summation 
makes the total number of summands into $6^\N$. Note also the important
point that the coupling parameter for background $\Phi$-field insertion is 
rescaled as $\lambda \rightarrow \lambda /2$ in the noncommutative case.

The two-loop Green functions Eq.(\ref{master1}) were originally sorted out 
by how the $\Phi$-field insertions are distributed among three internal
propagators. However, the result of the last section suggests that 
we ought to sort out the Green functions according to the  
way the $\Phi$-field insertions are distributed among three 
{\em boundaries}. Thus, we decompose the diagrams according to
the combinatorics
\bea
6^\N = (2 + 2 + 2)^\N =
\sum_{\sum{\N_i} = \N}
\frac{\N!}{\M_1!\M_2!\M_3!}
\left(\frac{\M_1!}{\N_2^+! \N_3^-!}\right)
\left(\frac{\M_2!}{\N_3^+! \N_1^-!}\right)
\left(\frac{\M_3!}{\N_1^+! \N_2^-!}\right).
\nonumber
\eea
In the summation, $\N_a^{\pm}$ are the number of momenta inserted on the
$(\pm)$-side of $a$-th internal propagator, and 
$\M_a = (\N_{a+1}^+ + \N_{a+2}^-)$ 
is the total number of momenta on the $a$-th boundary.

Specifically, the combinatoric factors organize 
the summation of the two-loop effective action in the following simple manner: 
\bea
\G &=& \sum_{\N} \frac{1}{\N!} \G_\N
\nonumber \\
\G_\N &=& \sum_{\{\M_a\}} \frac{\N!}{\M_1!\M_2!\M_3!} \G_{(\M_1, \M_2, \M_3)} 
\quad {\rm where} \quad (\M_1+\M_2+\M_3) = \N \nonumber
\eea
and
\bea
\G_{(\M_1, \M_2, \M_3)} &=& \prod_a \G_{\M_a}^{(a)} \nonumber
\\
\G^{(a)}_{\M_a} &=&  \sum_{\N^+_{a+1}, \N^-_{a+2}}
 \frac{\M_a!}{\N_{a+1}^+!\N_{a+2}^-!} 
\, \G^{(a)}_{\N_{a+1}^+, \N_{a+2}^-}
\qquad {\rm where} \qquad
(\N^+_{a+1} + \N^-_{a+2}) = {\rm M}_a.
\nonumber
\eea
As shown in section 3, $\G^{(a)}_{\N_{a+1}^+, \N_{a+2}^-}$
contains in its integrand, after Wick rotation back to Minkowski 
signature, terms of the combination: 
\bea
\left(- \frac{\l}{2}\right)^{\N_{a+1}^+ + \N_{a+2}^-}
\Big(\T_{a+1}\Big)^{\N_{a+1}^+} \Big(\T_{a+2}\Big)^{\N_{a+2}^-}
\widehat{J}_{(\N_{a+1}^+,\N_{a+2}^-)} \left(-\a_{a+1}, +\a_{a+2} \right).
\label{owlegg}
\eea
Recall that factors of $\T_{a+1}, \T_{a+2}$ originate from aforementioned 
rescaling of $\tau \rightarrow \T \tau$.
Taking the saddle-point Eq.(\ref{wow}) for $\T_a$'s and $\alpha_a$'s into 
Eq.(\ref{owlegg}), convoluting with the external $\Phi$-field insertions, and 
summing over $\N^+_{a+1}, \N^-_{a+2}$, 
we find factors of the following form:   
\bea
\vspace{-0.2cm}
&&\nonumber \\
&& \sum_{\N_{a+1}^+, \N_{a+2}^-} 
\left( - \frac{\lambda }{2} \right)^{{\rm M}_a}
\frac{(\T_{a+1})^{\N_{a+1}^+}} {\N_{a+1}^+!}
\frac{(\T_{a+2})^{\N_{a+2}^-}} {\N_{a+2}^-!}
\int \prod_{i=1}^{{\rm M}_a} {\d^d p_i \over (2 \pi)^d} 
(2 \pi)^d \delta\left(p_1 + \cdots + p_{\small {\rm M}_a} - k_a\right)
\nonumber \\
&& \hskip5cm \times \left[\widehat{J}_{(\N_{a+1}^+, \N_{a+2}^-)} \Big(-\a_{a+1}, +\a_{a+2}\Big)
\cdot \widetilde{\Phi}(p_1) \cdots \widetilde{\Phi}(p_{\small {\rm M}_a}) 
\right]
\nonumber \\
\vspace{-0.2cm}
\nonumber \\
&=& \sum_{\N_{a+1}^+, \N_{a+2}^-} 
\frac{\Big( -g|\a_{a+1}|\Big)^{\N_{a+1}^+}}{\N_{a+1}^+!} 
%\left( ig|\a_{a+1}|\right)^{\N_{a+1}^+}
\,\, \frac{\Big( -g|\a_{a+2}|\Big)^{\N_{a+2}^-}}{\N_{a+2}^-!}
%\left( ig|\a_{a+2}|\right)^{\N_{a+2}^-}
\int \prod_{i=1}^{{\rm M}_a} {\d^d p_i \over (2 \pi)^d}
(2 \pi)^d \delta( p_1 +\cdots + p_{\small {\rm M}_a} - k_a)
%\prod_{j=1}^{\N^+_{a+1}} {\d^d p^+_j \over (2 \pi)^d} 
\nonumber \\
&& \hskip5cm \times \left[
\widehat{J}_{(\N_{a+1}^+, N_{a+2}^-)}\Big(-\a_{a+1}, +\a_{a+2} \Big)
\cdot \widetilde{\Phi}(p_1)
\cdots 
%\widetilde{\Phi}(p^-_{\small \N^+_{a+1}})
%\widetilde{\Phi}(p^+_1) \cdots 
\widetilde{\Phi} (p_{\small {\rm M}_a}) \right]
\nonumber \\
\vspace{-0.2cm}
\nonumber \\
&:=& \widehat{W}_{(-\a_{a+1}, +\a_{a+2} )} [\Phi].
\label{owl}
\eea
At this stage, $\widehat{W}$ is merely a definition of weighted
resummation of the two-loop $\star_\N$-product kernel. Compared to the
one-loop $\star_\N$-product kernel, Eq.(\ref{owl}) indicates seemingly 
considerable differences. In particular, Eq.(\ref{owl}) depends on two
distinct vectors, $\alpha_{a+1}, \alpha_{a+2}$, and involves two 
exponentiations. Despite the differences, we will shortly 
show that $\widehat{W}$ is in fact a \underline{single} open Wilson line, 
but with a snapped contour!
%%%%%%%%%%%%%%%%%%%%%%%%%%%%%%%%%%%%%%%%%%%%%%%%%%%%%%%%%%%%%%%%%%%%%%%%%%%
\subsection{Interaction via Snapping of Open Wilson Lines}
%%%%%%%%%%%%%%%%%%%%%%%%%%%%%%%%%%%%%%%%%%%%%%%%%%%%%%%%%%%%%%%%%%%%%%%%%%%
How precisely is the product $\widehat{W}$ in Eq.(\ref{owl}) related to the 
scalar open Wilson line Eq.(\ref{scalarowl})? 
Recall that, from the definition Eq.(\ref{scalarowl}), shape of 
the open Wilson line contour was independent of the overall 
energy-momentum, $\sum_{i=1}^\N p_i = k$ or $\th \cdot 
\sum_{i=1}^\N p_i = \ell$, but depends only on how the the 
energy-momentum is distributed along the contour. For a {\sl straight} open 
Wilson line, Eq.(\ref{owldef}), the energy-momentum was uniformly distributed:
\bea
W_k[\Phi] = \int \d^d x {\cal P}_{\tau} 
\exp_\star \left( -g \vert \ell \vert \int\limits_0^1 \d \tau
\Phi(x+  \ell \tau) \right) \star e^{ik\cdot x}.
\nonumber
\eea
Taylor-expansion in powers of $(g \ell \Phi)$ yields:
\bea
W_k [\Phi] = \int \d^d x \Big[ && \hskip-0.3cm e^{ i k \cdot x}
\nonumber \\
&+& (-g\vert \ell \vert) \int\limits_0^1 \d \tau \Phi(x + \ell\tau) 
\star e^{i k \cdot x}
\nonumber \\
&+& (- g \vert \ell \vert )^2
\int\limits_0^1 \d \tau_1 \int\limits_{\tau_1}^1 \d \tau_2 
\Phi(x + \ell \tau_1) \star \Phi(x + \ell \tau_2) \star e^{ i k \cdot x}
\nonumber \\
&+& \cdots \Big]. 
\nonumber
\eea
Fourier-transform $\Phi$'s:
\bea
\Phi(x + \ell t) = \int {\d^d p \over (2 \pi)^d} \widetilde{\Phi}(p) \, 
{\bf T}_p \qquad {\rm where} \qquad {\bf T}_p := e^{ i p \cdot x},
\nonumber
\eea
where ${\bf T}_p$'s refer to elements of the `magnetic translation group' on the
noncommutative space, obeying the $\star$-product:
\bea
{\bf T}_p \star {\bf T}_q \,\, = \,\, e^{ {i \over 2} p \wedge q} \, {\bf T}_{p+q},
\nonumber
\eea
and then evaluate the parametric $\tau_1, \tau_2, \cdots$ integrals. 
The result is precisely the kernel $J_\N$ Eq.(\ref{starkernel}), in terms
of which the {\sl scalar} open Wilson lines, Eq.(\ref{owldef}) were defined. 
In fact, after Fourier-transforming
back to the configuration space, one finds
\bea
W_k [\Phi] = \sum_{\N=0}^\infty
\int \d^d x \, e^{ i k \cdot x} \, \sum_{\N=0}^\infty
(-g \ell)^\N {1 \over \N!} \left[\Phi(x) \cdots \Phi(x) \right]_{\star_\N} ,
\nonumber 
\eea
where the products involved, $\star_2, \star_3, \cdots$, are  the 
$\star_\N$-products. In fact, the $\star_\N$-product is identifiable
with the Parisi operator \cite{parisi}:
\bea
\left[\Phi \star_\N \Phi \right]_k 
:= {\cal P}_\tau \int \d^d x \Phi_1(x, k) \star \Phi_2 (x, k)
\cdots \Phi_\N (x, k) \star {\bf T}_k ,
\nonumber
\eea
where
\bea
\Phi_i (x, k) := \int\limits_0^1 \d \tau_i \, \Phi(x + \ell \tau_i).
\nonumber
\eea
The result reaffirms that the kernel $J_\N$ 
%Eq.(\ref{starkernel}) 
is Fourier-transform of the $\star_\N$-products.

Compare now the two-loop kernel $\hat{J}_{(\N_1^+,\N_2^-)}(-\a_1, \a_2)$ 
in Eq.(\ref{jhat}), or, equivalently, the resummed expression 
$\widehat{W}_{(-\alpha_{a+1}, \alpha_{a+2})}[\Phi]$ in Eq.(\ref{owl}) 
with the open Wilson line definition, Eq.(\ref{scalarowl}). Built upon 
the above discussions, one is led to suspect that $\widehat{J}$ arises in the 
Taylor-series expansion of a version of the open Wilson line, whose contour 
is along the vectors, $-\a_{a+1}$ and $\a_{a+2}$. 
First, the energy-momentum conservation across $a$-th boundary holds 
$\th\cdot(P_{a+1}^+ + P_{a+2}^-) = \ell_a = (\a_{a+2} - \a_{a+1})$. 
Second, the two vectors $\a_{a+1}, \a_{a+2}$ are not necessarily parallel,
so that the contour is snapped into a wedge-shape out of the straight 
contour $\ell_a$.
Third, the phase-factor Eq.(\ref{ff1}) contains 
the Moyal phase among the momenta within each segment 
as well as those coming across the two segments.
Fourth, the $\tau$-dependent exponential Eq.(\ref{ff2}) 
has the same structure as the straight open Wilson line. 
These observations assert that $\hat{J}_{(\N_{a+1}^+,\N_{a+2}^-)}
(-\a_{a+1}, \a_{a+2})$
together with the powers of $|{\a}|/2m$ in fact sum up to produce 
the snapped open Wilson line, which is now identified with 
$\widehat{W}_{(-\a_{a+1}, \a_{a+2})}[\Phi]$ in Eq.(\ref{owl})!

Care should be exercised in verifying this conclusion by an explicit 
computation. Recall that, for one-loop $\star_\N$-product kernel, 
the moduli parameters are periodic, $\tau_i \rightarrow \tau_i + 1$, reflecting
compactness of the one-loop vacuum Feynman diagram. Actually, as quoted in
footnote \ref{shift}, 
owing to the overall energy-momentum conservation, the kernel 
$J_\N$ for a straight Wilson line is invariant under an arbitrary translation 
of the moduli: $\tau_i \rightarrow \tau_i +$(constant).
This also implied that, despite its spacetime appearance as an open line, 
a straight open Wilson line treats all its marked points equally.
For a curved open Wilson line, the translation symmetry of moduli parameters
is lost, and there ought to be a physically preferred choice of the origin.
In view of the geometry of the interaction vertex in Fig. \ref{alpha}, for
the snapped open Wilson lines, it is natural to choose the point $P$ as 
the origin. 
%Even in this case, the orientation reversal remains as a valid
%symmetry: for instance, if we let $\tau$ run from $-1$ to $0$ on the $\a_1$ 
%side and from $0$ to $1$ on the $\a_2$ side, the result is exactly the same as 
%Eq.(\ref{jhat}). 

%%%%%%%%%%%%%%%%%%%%%%%%%%%%%%%%%%%%%%%%%%%%%%%%%%%%%%%%%%%%%%%%%%%%%%%%%%%%%
\section{Final Result}
%%%%%%%%%%%%%%%%%%%%%%%%%%%%%%%%%%%%%%%%%%%%%%%%%%%%%%%%%%%%%%%%%%%%%%%%%%%%%
Putting all together, the nonplanar part of the two-loop effective action is 
expressible as 
\bea
\label{finalfantasy}
\G\left[W[\Phi]\right] &=& {1 \over 24} \lambda^2 \hbar^2 
\int \frac{\d^d k_1}{(2\pi)^d} \cdots \frac{\d^d k_3}{(2\pi)^d}
(2 \pi)^d \delta^{(d)}(k_1+k_2+k_3) 
%\left(\frac{2m}{L}\right)^{d-3} 
%\left({\delta \T \over \T }\right)^3  
\nonumber \\
&\times& 
(4 \pi)^{-d} \Delta^{(d-3)/2} (\T)
\left({\Delta(\T) \over \Delta_F} \right)^{3/2}
\exp\Big(-m\left(|\a_1|+|\a_2|+|\a_3|\right)\Big) 
\nonumber \\
&\times& \exp\left(-\frac{i}{2} k_1\wedge k_2\right)
\widehat{W}_{(-\a_1, \a_2)} \widehat{W}_{(-\a_2,\a_3)}
\widehat{W}_{(-\a_3,\a_1)}.
\label{result1}
\eea
The overall coupling parameters originate from the combinatorics of
the two-loop {\sl vacuum} Feynman diagram. The second line is from the
saddle-point value of the moduli space integrals, where $\Delta(\T)$ and
$\Delta_F$ are computed and interpreted geometrically in Appendix B and C. 
In the third line, the phase-factor descends from the factorization, viz. 
the exponential in Eq.(\ref{decomp}), while the rest is from resummation of the 
background $\Phi$-field insertions. Finally, open Wilson lines are summed over 
the dipole sizes $\ell_a$'s, or, equivalently, over the boundary momenta
$k_a$'s, subject to the overall energy-momentum conservation.

Using the defining relation Eq.(\ref{elldef}), the phase-factor in
Eq.(\ref{result1}) is expressible as:
\bea
\exp\left( - {i \over 2} k_1 \wedge k_2 \right)
= \exp\left( {i \over 2} \alpha_1 \, {\tt V} \, \alpha_2 \right)
\exp\left( {i \over 2} \alpha_2 \, {\tt V} \, \alpha_3 \right)
\exp\left( {i \over 2} \alpha_3 \, {\tt V} \, \alpha_1 \right),
\nonumber
\eea
where ${\tt V}$ is a shorthand notation for wedge product with respect to
$(\theta^{-1})_{mn}$. Redefining the snapped open Wilson line so that each 
exponential
is absorbed into it:
\bea
\exp\left( {i \over 2} \alpha_1 \, {\tt V} \, \alpha_2 \right)
\widehat{W}_{(-\alpha_1, \alpha_2)}[\Phi]
\longrightarrow \widehat{W}_{(-\alpha_1, \alpha_2)} [\Phi],
\nonumber
\eea
the open Wilson lines in Eq.(\ref{result1}) are interpretable as matrices, 
whose row and column are indexed by $\{\alpha_a\}$, 
and their product corresponds to matrix multiplication! \footnote{ Note that the phase-factor absorbed vanishes 
identically for {\sl straight} open Wilson lines, and hence does not 
modify functional form of the result, Eq.(\ref{oneloopresult}).}  
Thus, while the open Wilson lines are naturally interpreted as composite
operators describing a `closed string', they seem to obey a noncommutative 
and associative algebra, rather than a commutative and non-associative 
one. 

Finally, introduce shorthand notations
\bea
 \left( -{\lambda \over 2} \right)^2 &:=& \lambda_{\rm c}
\nonumber \\
\int \prod_{a=1}^3 {\d^d k_a \over (2 \pi)^d}
(2 \pi)^d \delta (k_1 + k_2 + k_3) &:=& {\rm Tr}_{{\cal H}_{\rm dipole}}
\nonumber \\
(4 \pi)^{-d} \Delta(\T)^{(d-3)/ 2} \left( {\Delta (\T) \over \Delta_F} \right)^{3/2}
\prod_{a=1}^3 \exp \Big( - m \vert \alpha_a \vert \Big) &:=& {\bf K}_3
(\{\alpha_a\})
,
\nonumber
\eea 
the first being strongly reminiscent of the `soft-dilaton theorem' \cite{sdt}
in string theory. Then, Eq.(\ref{result1}) is re-expressible compactly as:
\bea
\Gamma_2 [W] = {\lambda_{\rm c} \over 3!}
{\rm Tr}_{{\cal H}_{\rm dipole}}
{\bf K}_3 (\{\alpha_a\}) \left( \widehat{W} \star \widehat{W} \star \widehat{W}
\right),
\nonumber
\eea
obtaining the proclaimed main result, Eq.(\ref{cubicint}). The $\star$'s
are to emphasize that the product involved are matrix multiplication, viz.
the newly emergent noncommutative geometry obeyed by the open Wilson lines. 

\begin{figure}[htb]
\centerline{\epsfxsize=9cm\epsfbox{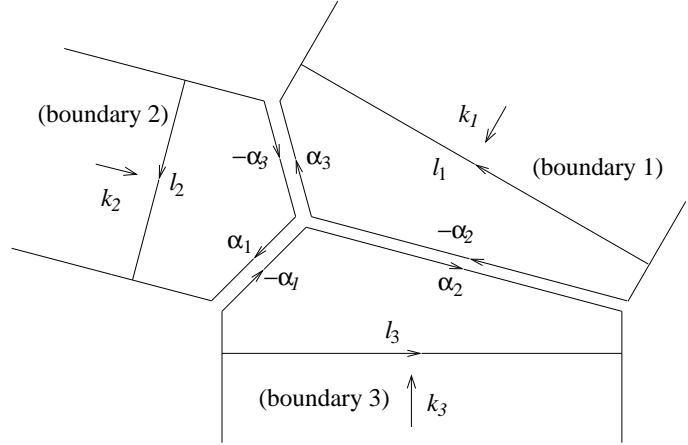}}
\caption{\sl Spacetime view of the cubic interaction among the noncommutative 
dipoles created by the open Wilson line operators.} 
\label{alpha2} 
\end{figure}

Pictorially, the cubic
interaction among the `closed string' states created by the open Wilson line
operators is depicted in Fig.\ref{alpha2}. The interaction goes as follows.
At asymptotic infinity, three {\sl straight} noncommutative dipoles, viz. 
free `closed strings', approach one another. Near the interaction point 
$P$, each dipole snaps its contour dynamically so that the interaction with 
adjacent dipoles is purely local and maximally overlapping. 

Note that the effective action, once expressed in terms of the snapped
open Wilson lines, is remarkably similar to the half-string 
formulaton \cite{half1, half2} of Witten's open string field theory 
\cite{wosft}, though the noncommutative dipoles are interpretable most 
naturally as `closed strings'. 
A minor difference is that location of $P$ is determined {\sl dynamically} 
by the energy-momenta of the asymptotic dipoles, while, in Witten's open 
string field theory, it is fixed at the midpoint once and for all, breaking
the worldsheet reparametrization invariance.   
We trust that the similarity is far more than a mere coincidence, 
and further elaboration of the relation will be reported elsewhere.

Finally, in light of the ubiquity of the open Wilson lines, precisely the 
same result and physical interpretation can be drawn for noncommutative 
field theories, including noncommutative gauge theories \footnote{A step
toward this direction is made in \cite{armonilopez}.} and those with 
supersymmetries. Explicit verification for such theories are straightforward, 
and will be reported in forthcoming papers.

%%%%%%%%%%%%%%%%%%%%%%%%%%%%%%%%%%%%%%%%%%%%%%%%%%%%%%%%%%%%%%%%%%%%%
\subsection*{Acknowledgements}
%%%%%%%%%%%%%%%%%%%%%%%%%%%%%%%%%%%%%%%%%%%%%%%%%%%%%%%%%%%%%%%%%%%%%
We acknowledge enlightening discussions with Costas Bachas, Hong Liu, 
Peter Mayr, Jeremy Michelson, Ashoke Sen, Erik Verlinde, Herman Verlinde, 
and Jung-Tay Yee. SJR thanks warm
hospitality of CERN Theory Division, Spinoza Institute at Utrecht University, 
Summer Institute 2001 at Yamanashi-Japan, and Ecole de Physiqu\'e Les Houches, 
where substantial parts of the present work were accomplished. 

\section*{Appendix}
\appendix
%%%%%%%%%%%%%%%%%%%%%%%%%%%%%%%%%%%%%%%%%%%%%%%%%%%%%%%%%%%%%%%%%%%%%%%%%%%
\section{Two-Loop Feynman Diagrammatics}
%%%%%%%%%%%%%%%%%%%%%%%%%%%%%%%%%%%%%%%%%%%%%%%%%%%%%%%%%%%%%%%%%%%%%%%%%%%
In this appendix, we will elaborate derivation of Eqs.(\ref{GPN}, 
\ref{master1}) via the noncommutative Feynman diagrammatics.
The noncommutative Feynman rules of
$\lambda[\Phi^3]_\star$-theory are summarized, 
in the background field method,  by the following generating 
functional:
\bea
Z[\Phi_0] = Z_0[\Phi_0]
\int {\cal D} \varphi \exp 
\left( - \int \d^d x \left[ {1 \over 2} \varphi (x) 
( - \partial_x^2 +m^2 + \lambda \Phi_0(x)) \varphi(x) 
+ {\lambda \over 3!} \varphi^3(x) \right]_\star \right).
\label{rule}
\eea
Here, $\Phi_0(x)$ and $\varphi(x)$ refer to the background and
the fluctuation parts of the scalar field, $\Phi$, respectively.
In the notations and conventions explained in the previous subsection, 
the two-loop Feynman integral of the N-point, one-particle-irreducible 
Green function is given, after Wick rotation to the Euclidean space, by
\begin{eqnarray}
\G_\N & = & (-\lambda)^{{\rm N} +2 } C_{\{\N\}}
\int {\d q_1 \over (2 \pi)^d} { \d q_2 \over (2 \pi)^d} 
\prod_{a=1}^3 \left( 
\prod_{i=0}^{{\rm N}_a} \frac{1}
{\left( q_a + \sum_{j=1}^i p_j^{(a)} \right)^2 + m^2} 
\right)
\label{feynstar} \\
&\times& \!\! \exp \left[  \frac{i}{2} \left( q_1 \!\wedge\! q_2 
- (q_1 + P_1) \!\wedge\! (q_2 + P_2) \right) \right] 
\exp \left[ - \frac{i}{2}\sum_{i=1}^{{\rm N}_a} \Big( q_a + 
\sum_{j=1}^{i-1} p_j^{(a)} \Big) \!\wedge\! p_i^{(a)} \nu_i^{(a)}  \right]. 
\nonumber
\end{eqnarray}
Here, $q_a$'s denote the momentum flowing through the $a$-th internal
propagator ($a=1,2,3$), and are subject to the momentum conservation, 
$q_1 + q_2 +q_3 = 0$. As compared to the commutative counterpart,
the integrand in Eq.(\ref{feynstar}) 
is modified by the two phase-factors arising from the Moyal's
product in Eq.(\ref{rule}). The first phase-factor originates from Moyal's
product present in the two cubic interaction vertices. The second
originates from insertion of each external line, and depends on $\nu$ 
\footnote{The convention we adopt for the Feynman rules is that all 
phase-factors reside at the interaction vertices. In the convention adopted 
in the worldline formulation \cite{master}, an overall phase is assigned 
for each diagram and, at the same time, for nonplanar an appropriate extra 
phase-factor is attached for each nonplanar crossing. It can be shown 
straightforwardly
that the two phase-factor conventions produce identical results.}. 

To facilitate the integration, introduce the Feynman-Schwinger parametric
representation to each of the total (N+3) propagators: 
\bea
\left[ \frac{1}{ \left( q_a 
+ \sum_{j=1}^i p_j^{(a)} \right)^2 + m^2 } \right]
= \int\limits_0^\infty  \d s_i^{(a)} 
\exp \left[ -s^{(a)}_i \left( 
\left( q_a + \sum_{j=1}^i p_j^{(a)} \right)^2 + m^2 \right) \right] ~ .
\nonumber
\eea
Decompose $s_i^{(a)}$'s into N `insertion position moduli' parameters 
$\tau_i$'s and the `vacuum moduli' $\T_a$'s: 
\bea
\T_a \, &:=& \sum_{j=0}^{{\rm N}_a} s_j^{(a)} 
 = \t_0^{(a)} \quad \qquad (a = 1, 2, 3)
\nonumber \\
\t_i^{(a)} &:=& \sum_{j=i}^{{\rm N}_a} s^{(a)}_j
\qquad \qquad \quad (i = 1,2, \cdots, \N_a).
\label{moduli}
\eea
The Jacobian in changing variables 
from $\left(s_0^{(a)} , \cdots , s_{{\rm N}_a}^{(a)} \right)$ 
to $ \left(\T_a, \tau_1^{(a)} , \cdots , \tau_{{\rm N}_a}^{(a)} \right)$ is 
evidently unity. 
In terms of the newly introduced moduli parameters, Eq.(\ref{moduli}), 
the string of propagators along each of the three vacuum internal lines is
expressible as:
\begin{eqnarray}
&& \prod_{a=1}^3  \prod_{i=0}^{{\rm N}_a} 
\left( \frac{1}{\left( q_a + \sum_{j=1}^i p_j^{(a)} \right)^2 + m^2 } 
\right) = \prod_{a=1}^3 \int\limits_0^{\infty} \d \T_a 
 \int \d \tau^{(a)}_1 \cdots d\tau_{{\rm N}_a}^{(a)} e^{- m^2 \T_a}
\label{elf} \\
 &\times& \exp \left[ 
- \T_a (q_a)^2  - 2 q_a 
 \cdot \sum_{ i =1}^{{\rm N}_a} p^{(a)}_i \t^{(a)}_i 
  - \sum_{i=1}^{{\rm N}_a} p_i^{(a)} \cdot p_i^{(a)} \t^{(a)}_i - 
2 \sum_{i<j}^{{\rm N}_a} p^{(a)}_i \cdot p^{(a)}_j \tau^{(a)}_j \right]. 
\nonumber
\end{eqnarray}
Here, the $\tau^{(a)}$-integrations are defined over $0 \le \tau_{\N_a}^{(a)} 
\le \cdots \le \tau_{1}^{(a)} \le \T_a \le \infty $. 
One can readily show that, once summed over all possible permutations of 
external insertions along each vacuum internal line, all the 
$\tau^{(a)}_i$-integrations are extended over $[ 0, \T_a )$. 

Next, we evaluate the integrals over the loop momenta $q^{(1)}$ and
$q^{(2)}$. Recall first how the integrals are evaluated in the
commutative case, where the two exponential functions in Eq.(\ref{feynstar})
involving noncommutative phase-factors are absent. 
Collecting the $q^{(a)}$-dependent part in the exponent of Eq.(\ref{elf}):
\bea
\label{kall}
 -\T_1 (q_1)^2 -\T_2 (q_2)^2 -\T_3 ( q_1 + q_2 )^2
 -2q_1 \cdot A_1 -2q_2 \cdot A_2 +2( q_1 + q_2 ) \cdot A_3 ~ ,
\eea
where $A_a := \left( \sum p_i\t_i \right)_a$, and expressing the quadratic
polynomial into a complete-square function, the integrals over $q_1, q_2$
can be performed successively as Gaussian integrals. One can express the 
result \footnote{
The result is identical to those presented in \cite{rolandsato}, where
the relationship between the worldline propagators, $G_{ab}$,
and the string theory worldsheet propagators are elaborated in detail.}
compactly as \footnote{
In obtaining the result, the following 
idendities resulting from momentum conservation have been used:
\bea
- \left[
\sum_i p_i^2 \t_i + 2 \sum_{i<j} p_i\cdot p_j \t_j
\right]_a 
&=& 
\left[
\sum_{i<j} p_i\cdot p_j \t_{ij}
\right]_a + A_a (P_{a+1} + P_{a+2} )
\nonumber
\\
- \left[
\sum_i p_i^2 \t_i^2 + 2 \sum_{i<j} p_i\cdot p_j \t_i\t_j
\right]_a 
&=& 
\left[
\sum_{i<j} p_i\cdot p_j \t_{ij}^2 
\right]_a  + \left(\sum p_i\t_i^2 \right)_a (P_{a+1} + P_{a+2} ).
\nonumber
\eea}
:
\bea
\Delta^{d \over 2} (\T) \left(\prod_{a=1}^3  e^{-m^2 \T_a}\right)
\exp \left[ \half \sum_{i=1}^{\N_a} \sum_{j=1}^{\N_b} 
G_{ab}^{(0)} \left(\tau^{(a)}_i, \tau^{(b)}_j \right) 
p^{(a)}_i \cdot p^{(b)}_j \right], 
\label{commresult}
\eea
where
\bea
G_{aa}^{(0)} (x,y) &=& |x-y| -\Delta \left(\T_{a+1}+\T_{a+2} \right)(x-y)^2 ~ ,
\nonumber
\\
G_{aa+1}^{(0)} (x,y) &=& x+y - \Delta \left(x^2 \T_{a+1} + y^2 \T_a + 
  (x-y)^2 \T_{a+2} \right) ~ ,
\nonumber
\eea
and account for, in Eq.(\ref{master1}), 
the $\T_a$-dependent weight functions in the first line,
and the first exponential in the second line. 

We now turn to the phase-factors arising from the noncommutative 
Feynman rules. They arise from the two 
cubic interaction vertices, viz. the phase-factor of the first
exponential in Eq.(\ref{feynstar}): 
\be 
\frac{i}{2} q_1 \wedge q_2 
- \frac{i}{2} (q_1 + P_1) \wedge (q_2 + P_2) 
= -\frac{i}{2} q_1 \wedge P_2 + \frac{i}{2} q_2 \wedge P_1 
-\frac{i}{2} P_1 \wedge P_2,
\label{phase1}
\ee
and from the external insertions, viz. the phase-factor in the second
exponential in Eq.(\ref{feynstar}):
\bea
&&
- \frac{i}{2}\sum_{i=1}^{N} \left( q_a + 
\sum_{j=1}^{i-1} p_j^{(a)} \right)
\wedge p_i^{(a)} \nu_i^{(a)} 
\label{phase2} \\
&=&
- \frac{i}{2} q_a \wedge (P^+_a -P^-_a ) - \frac{i}{2} P^-_a 
 \wedge P^+_a
-\frac{i}{4} \sum_{i<j}^{{\rm N}_a} p^{(a)}_{i}
\wedge p^{(a)}_{j} \left( \nu^{(a)}_{i}+\nu^{(a)}_{j} \right)
\epsilon \left(\tau^{(a)}_{ij} \right)
\nonumber
\eea
along each $a$-th internal line.  First, note that the 
moduli-independent phase-factors from Eqs.(\ref{phase1},\ref{phase2})
combine and yield the exponential with moduli-independent
phase-factors (via momentum conservation
$P_1 + P_2 + P_3 = 0$)
\begin{equation}
\exp  \left[ - \frac{i}{2} \sum_{a=1}^3 (
 P_a^- \wedge P_a^+ ) - \frac{i}{2} P_1 \wedge P_2 \right]
= \exp \left[ - \frac{i}{2} \sum_{a=1}^3 \left(
 P_a^- \wedge P_a^+ + \frac{1}{3} P_a \wedge P_{a+1}
\right) \right] ~ ,
\end{equation}
and an exponential with $\epsilon \left(\tau_{ij}^{(a)} \right)$-dependent 
phase-factors
\begin{equation}
\exp\left[\, -\frac{i}{4} \sum_{i<j}^{{\rm N}_a} p^{(a)}_{i}
\wedge p^{(a)}_{j} \left(\nu^{(a)}_{i}+\nu^{(a)}_{j} \right)
\epsilon \left(\tau^{(a)}_{ij} \right)\,\right]. 
\end{equation}
They account for the two exponentials in Eq.(\ref{master1}): 
$\Xi_1, \Xi_2$, respectively. 

Note that, in Eq.(\ref{phase1}), terms quadratic in $q^{(a)}$ cancel
each other. This is inherited to the fact that, as mentioned in 
footnote 3, 
the two-loop vacuum diagram under consideration descends, in the Seiberg-Witten
limit of bosonic string theory, from disk worldsheet with two holes or, 
equivalently, from sphere worldsheet with three holes, and hence referred it
as {\sl planar} vacuum diagram. If one of the propagators emanating from the
first interaction vertex were  twisted before joining the second vertex, 
the resulting vacuum diagram is {\sl nonplanar}, for which the sign of the 
second term in the first expression of Eq.(\ref{phase1}) is reversed.  
Thus, for nonplanar diagrams, terms quadratic in $q^{(a)}$'s, $ i q^{(1)} 
\wedge q^{(2)}$, are produced \footnote{In \cite{master}, it was shown that
the nonplanar diagram contribution to the ${\rm N}$-point Green function 
can be related to the planar diagram contribution via insertion of a  
pseudo-differential operator, 
$\exp ( - i \partial_z \wedge \partial_w ) |_{z = w = x_2}$, 
where $x_2$ refers to the spacetime position of the second interaction
vertex. }. 

Cancellation of terms quadratic in $q^{(a)}$'s puts the rest of 
the computation straightforward. Collect terms linear in $q_a$:
\begin{eqnarray}
-\frac{i}{2} \Big[ q_1 \!\wedge\!  (P_1^+ - P_1^-) 
                   + q_2 \!\wedge\! (P_2^+ - P_2^-) 
                   - (q_1+q_2) \!\wedge\! (P_3^+ - P_3^-) 
+ q_1 \!\wedge\! P_2 - q_2 \!\wedge\! P_1 \Big], 
\label{qphase}
\end{eqnarray}
add this to Eq.(\ref{kall}), and perform the integrations over 
$q_1$ and $q_2$.  As the quadratic parts in the exponent
are not modified, compared to the commutative case Eq.(\ref{commresult}), 
the integration results in two extra phase-factors: 
a $\wedge$-product originating from cross terms between 
Eq.(\ref{kall}) and Eq.(\ref{qphase}), and a $\circ$-product 
originating from the square of Eq.(\ref{qphase}).
The two extra phase-factors are further simplifiable in terms of the 
{\sl worldsheet boundary} momenta $k_a$:
\bea
\exp \Delta \sum_{a=1}^3 \left( i \T_a k_a \wedge 
(A_{a+2} - A_{a+1}) 
-\frac{1}{4} \T_a k_a \circ k_a \right) ~ .
\nonumber
\eea
This accounts for, in Eq.(\ref{master1}), the second exponential in
the second line and the last phase-factor, $\Xi_3$.

%%%%%%%%%%%%%%%%%%%%%%%%%%%%%%%%%%%%%%%%%%%%%%%%%%%%%%%%%%%%%%%%%%%%%%%%%
\section{Saddle-Point Analysis}
%%%%%%%%%%%%%%%%%%%%%%%%%%%%%%%%%%%%%%%%%%%%%%%%%%%%%%%%%%%%%%%%%%%%%%%%%
The saddle point condition should determine the
saddle point value of $\T_a$'s in terms of $l_a$'s.  
In this Appendix, we sketch a method to achieve that
task and determine $L$ as a consequence.  

The saddle point condition demands that the three vectors
$(\a_a/t_a) = t_{a+1} l_{a+1} - t_{a+2} l_{a+2}$ form an equilateral 
triangle.
The length of the vectors is defined to be $L$. 
The products of $l_a = \a_{a+2} - \a_{a+1}$ 
thus take the simple form:
\bea
\label{l1}
2l_a\cdot l_{a+1} &=& (1-2t_{a+2}(t_1+t_2+t_3)) L^2 , \\
l_{a}^2 &=& (t_{a+1}^2+t_{a+2}^2 + t_{a+1}t_{a+2} )L^2 ,
\label{l2}
\eea
We have used the normalization condition $(t_1t_2 + t_2t_3 +t_3t_1 = 1)$ 
to simplify Eq.(\ref{l1}). One way to solve these equations is 
as follows. First, we rewrite Eq.(\ref{l1}) as
\bea
\label{l3}
t_{a+2} = \frac{L^2 - 2 l_a \cdot l_{a+1}}{2L^2 (t_1+t_2+t_3)}.
\eea
We then sum up Eq.(\ref{l2}), and find
\be
\label{l4}
2 L^2 (t_1+t_2+t_3)^2 = l_1^2 + l_2^2 + l_3^2 +3L^2.
\ee
We can use Eq.(\ref{l4}) to eliminate $(t_1+t_2+t_3)$ from Eq.(\ref{l3}),
\be
t_{a+2} = \frac{L^2 - 2 l_a \cdot l_{a+1}}
{\sqrt{2L^2(l_1^2 + l_2^2 + l_3^2 +3L^2)}}.
\label{expt}
\ee
Plugging the $\{t_a\}$ of Eq.(\ref{expt}) into the normalization 
condition and further using momentum conservation determines
$L$: 
\bea
3 L^4 &=& 4\{ 
(l_1\cdot l_2)(l_2\cdot l_3) + 
(l_2\cdot l_3)(l_3\cdot l_1) + 
(l_3\cdot l_1)(l_1\cdot l_2) \}
\nonumber \\
&=& -(l_1^4 + l_2^4 +l_3^4) +2 ( l_1^2l_2^2 + l_2^2 l_3^2 + l_3^2l_1^2).
\nonumber
\eea
Comparing the last line with Heron's formula \cite{heron}
for the area of a triangle, 
we find that 
\be
\frac{\sqrt{3}}{4} L^2 = {\rm Area}(l_1,l_2,l_3).
\label{expl}
\ee
Note that the left-hand-side is the area of the {\sl equilateral} triangle
whose sides have the length $L$, while the right-hand-side is the area of 
the triangle formed out of $\{l_a\}$.  The expressions Eq.(\ref{expt}) 
and Eq.(\ref{expl}) 
explicitly determine the saddle point $t_a$ (and thus $T_a$) in 
terms of $l_a$'s, from which we can compute every quantities
of interests at the saddle point.  
 
If we are interested in deriving Eq.(\ref{expl}) only, 
there is a simpler way, though 
it does not give the value of $t_a$.  Recall that $\{ \a_a \}$ divide 
the triangle into three disjoint pieces. The area of the triangle is 
then given by
\bea
A &=& \half |\a_1||\a_2| \sin(2\pi/3)
+ \half |\a_2||\a_3| \sin(2\pi/3)
+ \half |\a_3||\a_1| \sin(2\pi/3)
\\
&=& \frac{\sqrt{3}}{4} L^2 (t_1t_2+t_2t_3+t_3t_1) = \frac{\sqrt{3}}{4} 
  L^2 ~ , 
\nonumber
\eea
where we have used the normailzation condition, obtaining the final result. 

%%%%%%%%%%%%%%%%%%%%%%%%%%%%%%%%%%%%%%%%%%%%%%%%%%%%%%%%%%%%%%%%%%%%%%%%%%%%%
\section{Computation of $(\Delta_F)^{-3/2}$}
%%%%%%%%%%%%%%%%%%%%%%%%%%%%%%%%%%%%%%%%%%%%%%%%%%%%%%%%%%%%%%%%%%%%%%%%%%%%%
Recall that the form of the exponent is given 
by 
\bea
F = -m^2 (\T_1 + \T_2 + \T_3) - \frac{1}{4} 
\Delta(\T) (\T_1l_1^2 + \T_2l_2^2+ \T_3l_3^2) ~ .
\nonumber
\eea
The saddle point conditions are
\bea
\frac{\partial F}{\partial \T_a} = -m^2 + 
\frac{\Delta^2}{4} (\T_{a+1}l_{a+1} - \T_{a+2}l_{a+2})^2 = 0 ~ .
\nonumber
\eea
The second derivatives of $F$ at the saddle point turn out to be
\bea
\frac{\partial^2 F}{\partial \T_a^2} &=& -4 (t_{a+1}+t_{a+2}) 
  \frac{m^3}{L} ~ ,
\nonumber
\\
\frac{\partial^2 F}{\partial \T_a \partial \T_{a+1}} 
&=& - 2 t_{a+2} \frac{m^3}{L} ~ .
\nonumber
\eea
The width of the saddle is thus given by
\bea
(\Delta_F )^{-3/2}
&=& 
(2\pi)^{3/2}
\det\left(\frac{\partial^2 F}
{\partial \T_a \partial \T_b} \right)^{-1/2}
\nonumber \\
&=& \left( \frac{\pi^3 L^3}{6m^9} \right)^{1/2}
  (t_1 +t_2+t_3)^{-1/2} ~ .
\nonumber
\eea
It is worth noting that the dimensionless width is expressible as
\bea
\left(\frac{32\pi^3}{3}\right)^{-1/2}
\left( {\Delta (\T) \over \Delta_F }\right)^{3/2} 
= (mL)^{-3/2}(t_1+t_2+t_3)^{-1/2} 
= (m^3 L^2 (|\a_1|+|\a_2|+|\a_3|))^{-1/2} .
\nonumber
\eea

\end{document}